# Allocating Duplicate Copies for IoT Data in Cloud Computing Based on Harmony Search Algorithm

Younes Jahandideh[1], A. Mirzaei[1*]

[1,] Department of Computer Engineering, Ardabil Branch, Islamic Azad University, Ardabil, Iran

**ABSTRACT**

The Internet of things (IoT) generates a plethora of data nowadays, and cloud computing has been introduced as an efficient solution to IoT data management. A cloud resource administrator usually adopts the replication strategy to guarantee the reliability of IoT data. This mechanism can significantly reduce data access time, and evidently, more replicas of data increase the data storage cost. Furthermore, the process of selecting mini clouds for replica allocation and sorting replicas in mini clouds is considered an NP-hard problem. Therefore, this paper proposes an approach based on the harmony search (HS) algorithm to allocate replicas to the IoT data in the cloud computing environment in order to mitigate the data access cost. The HS algorithm was employed in the proposed approach to determine the best location for data replication in the cloud computing environment. According to the implementation results, the proposed approach outperformed the other methods and managed to significantly decrease data access time and delay as well as energy consumption.

## 1. INTRODUCTION

As two entirely distinct technologies, IoT and cloud computing have complementary features that evolve rapidly. In fact, the IoT benefits from the unlimited storage and processing power of clouds, and in return, cloud computing can take advantage of the IoT features to provide more distributed and dynamic communications with real-world objects. Cloud computing focuses mainly on the development of powerful data centers to provide end-users with dynamic and flexible services [1]. Therefore, when the IoT transmits the data sensed by wireless sensors to the cloud for storage, multiple replicas of the transmitted data are stored in the cloud so that the data can be sent from the closer server to the end-user upon request. The replicas of IoT data are stored in the cloud computing environment to retain data and prevent data deletion and loss in the cloud and also accelerate the end-users access to data. Since heuristic methods operate slowly in selecting data centers for data replication, meta-heuristic methods can be a better alternative choice [2]. Accordingly, this paper employed the HS algorithm to propose an approach to the allocation of IoT data replicas in the cloud computing environment. The HS algorithm was used because the proposed approach can properly benefit from its advantages in both local and global search procedures, thus covering the weaknesses of other existing methods. Inspired by music, the HS algorithm aims to reach the best solution through coordination and harmony. The attempt at finding harmony and coordination in music resembles the discovery of optimal conditions in an optimization process.

This study integrated access frequency, data dependency, data center storage capacity, and dataset size in order to develop data replication mechanisms for each dataset. In the proposed approach, the data storage space is divided into two sections, and the data are then categorized in three classes. A data replication algorithm is also created at the development time with a different determinant level of data dependency, access frequency, data center data storage capacity, and dataset size. According to the case study, the proposed approach managed to significantly mitigate the total costs of data storage and information transfer for software applications. In this approach, an algorithm was proposed to obtain and allocate data replicas hierarchically in a wireless mesh network by using the graph division solution.

This paper consists of the following sections. Related works and motivation are reviewed in Section 2. We describe the system model and the proposed HS algorithm in Section 3. Section 4 discusses mapping of HS algorithm onto the replica allocation problem. The results of implementing the proposed approach are evaluated in Section 5. Finally, Section 6 presents the conclusion and future works.

## 2. RELATED WORKS AND MOTIVATION

In cloud computing environments, replication is performed to achieve high resource availability. Resources are copied dynamically to mitigate the total cost of retaining the compatibility of replicated and copied files. Different strategies have been proposed to manage and allocate a plethora of IoT datasets in the cloud environment. In [3], QDR: A QoS-Aware Data Replication Algorithm for Data Grids Considering Security Factors proposed a novel location-allocation algorithm by considering the availability of data, the number of requested replicas, and the size of the replicated data. The proposed algorithm aimed to enhance network efficiency and increase the accessibility rate of files and data. In [4], an algorithm named LEARN (LatEncy Aware Replica placemeNt) was proposed to allocate the replicas of data in virtual disks of a cloud. In this algorithm, replicas can be placed in virtual disks at the beginning of storage without migration. LEARN decreases the mean network delay by allocating the replicas of data to virtual disks. In [5], an efficient algorithm was proposed to minimize the time required to copy data to a new location based on the migration algorithm. The proposed algorithm aimed to mitigate the void network congestion and guarantee appropriate accessibility of data. In [6], an adaptive sorting method was proposed to consider the communicative performance between different computational nodes in a cloud. In [7], a heuristic algorithm was proposed to transfer network administration to the cloud. The proposed metaheuristic algorithm is actually a prototype implemented in the delivery, updating, and replication processes. In [8], an efficient smart control plan was proposed for the allocation of replicas in the cloud by using cryptography for replication. This algorithm supports the dynamic operations of block data and third-party public validation to provide high security against data forgery and replacement. In [9], a replacement algorithm was proposed for data replicas based on the servers existing in a network in order to improve the replicated service and decrease user costs. In [10], a replication algorithm was proposed to reduce the access delay and expand the network bandwidth. In [11], a new data replication strategy was proposed to mitigate the costs of data storage and information transfer for applications. In [12], data replication schemes were reviewed achieving higher performance for data-intensive applications. Also, some researchers addressed main critical challenges of this criterion, such as reliability [13] availability [14], security [15], bandwidth [16], and response time [17] of data access. However, cloud computing has security challenges, including vulnerability for clients and association acknowledgment, that delay the rapid adoption of computing models. In [18], the authors reviewed different machine learning algorithms to deal with the cloud security challenges. Many supervised learning algorithms such as MLP [19], CNN [20], LSTM [21,22], and unsupervised learning algorithms such as LDA [23], K-Harmonic Means [24], DBN [25] were presented for analyzing security threats. Machine learning has also been used to achieve the optimal network configuration considering resource allocation and scheduling [26], [27]. In [28], a semi-supervised learning algorithm (Q-learning) was used to dynamically adjust the appropriate number of resources for cloud computing, which has a better performance in resource management than other methods.

In brief, although different studies have been conducted on allocation of replicas in cloud computing environments in recent years, there is still scant literature on the reduction of cost, delay, and energy consumption through metaheuristic algorithms. This paper proposes a strategy for the allocation of replicas based on the HS algorithm in cloud computing environments in order to mitigate the abovementioned parameters. The above-mentioned issues have been exacerbated during the development of IoT, where few of the current systems can reduce the deployment costs while benefiting from the heterogeneous wireless devices. Different from home IoT scenarios, system infrastructure settings in industrial applications are normally infeasible to be fundamentally changed unless a huge retrofitting. Therefore, how to scale the current IoT by using heterogeneous IoT devices becomes a challenging issue. Facing these challenges, we take advantages of the harmony search algorithm to allocate replicas to the IoT data in the cloud computing environment in order to mitigate the data access cost in order to help improve the scalability of the current IoT network.

Despite the existing works, the proposed algorithm applies a hybrid consisting of K-means and P-center algorithms based on the weighted mean of response time to develop data replication mechanisms for each dataset.

## 3. SYSTEM MODEL AND HARMONY SEARCH ALGORITHM

### 3.1 System Model

In this replica allocation problem, a sensor's information is assumed to be stored in different locations. Therefore, the number of replicas is different for each piece of information. Figure 1 illustrates the general system model in the proposed approach. In this model, objects access the cloud environment through gateways for processing and storage. Every gateway has two parameters, *i.e.* information reading delay and data waiting time. The cloud environment contains a few mini clouds, each including a specific number of servers (with a specific capacity, reading delay, and writing delay). Hence, according to the proposed method, the data are first delivered to the first cloud for replication and then sent to mini clouds for allocation.

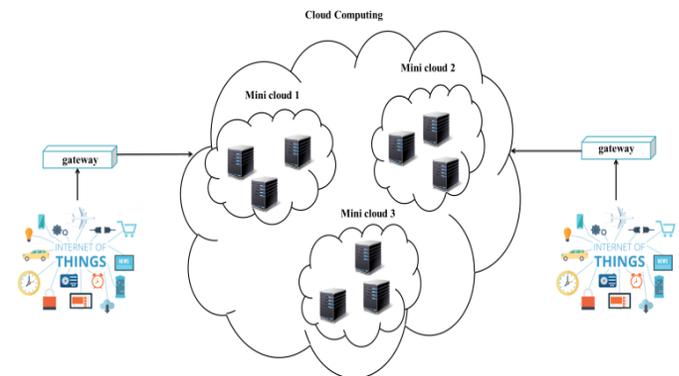

**Figure 1.** The general cloud computing environment model.

The process of finding the best state (selecting mini clouds) for data replication and sorting replicas in mini clouds is considered an NP-hard problem. The proposed approach employs the HS algorithm to perform this process and achieve our most important goal, *i.e.,* faster data accessibility considering the cost function. Table 1 illustrates the parameters of the cost function.

Table 1. Parameters of the cost function.

| Parameter | Description |
|---|---|
| n | Number of mini clouds |
| $T_{dj}$ | Waiting time of $D_d$ in $G_j$ |
| $B_{jc}$ | Data transfer rate from gateway $j$ to cloud $c$ |
| $R_j$ | Reading delay in gateway $j$ |
| $W_c$ | Writing delay in cloud $c$ |
| $L_d$ | Size of $D_d$ in $G_j$ |
| $c'$ | A series of clouds as many as the number of replicas required for data except for one |
| $T_{dc}$ | Waiting time of $D_d$ in $C_c$ |
| $B'_{cc'}$ | Data transfer rate from $c$ to $c'$ |
| $R_c$ | Reading delay in $c$ |
| $W_{c'}$ | Writing delay in $c'$ |
| $D_d$ | Data waiting time |
| $G_j$ | Gateway $j$ |
| $C_c$ | Cloud $c$ |

### 3.2 Harmony Search Algorithm

The harmony search (HS) algorithm [29] is inspired by the process of playing music and trying to find an extraordinary harmony accordingly. This process aims to find an appealing and pleasurable harmony (a complete state of harmonies) based on the harmony aesthetics. In different musical instruments, a selected note determines the aesthetics of the resultant harmony. In a musical process, a player starts playing music and searching for better harmony. Considering their previous works, players try to play the best of their previous harmonies or improve them. Players can also improvise a new piece of music about which they have no prior experience. In the process of playing music with no prior experience, players can play every note in its authorized interval so that it can produce a harmony vector along with the other notes. If desirable, the resultant harmony is stored in the player's mind, increasing the probability of producing better harmonies in the next practices and plays. The most important component of HS is the harmony memory that includes a specific number of harmony vectors, each of which results from the act of playing. For instance, consider jazz music which consists of three musical instruments. If each of these instruments selects an authorized note, the output will be a vector of harmonies. If the quality and aesthetics of this vector outdo the worst previous experience, the vector will replace it in the harmony memory. The act of playing is repeated in order to find an extraordinary harmony in the memory. In the actual optimization, every player and every acoustic note are replaced by a variable and the corresponding value, respectively. As discussed earlier, the player has three options for playing a note: playing one of the notes existing in his memory, playing a note close to one of the notes existing in his memory, or playing a totally random note in an authorized interval. In the HS algorithm, the value of every variable is determined through one of the following three rules to generate a new vector of variables (harmony vector):

A) Selecting a value existing in the harmony memory

B) Selecting a value in proximity to one of the values existing in the harmony memory

C) Selecting a random value for the variable from the authorized interval (outside the harmony memory)

## 4. MAPPING THE HS ALGORITHM ONTO THE REPLICA ALLOCATION PROBLEM

In the HS algorithm, every harmony memory includes a solution that covers the problem space and indicates a solution to the transmission of data replication on storage resources in the cloud environment. After each replication, the existing solution is considered a harmony and its cost function is calculated. The HS algorithm can be conveniently employed for calculating and applying appropriate values for relevant parameters, thus increasing the probability of achieving optimal solutions. The HS algorithm operates based on a schedule to evade local optimization. In fact, it starts searching for optimal solutions and replicating them until an optimal solution is obtained. After the initial population is created in the HS algorithm, a cost function is employed to evaluate the appropriateness or inappropriateness of a solution. Figure 2 shows the flowchart for the proposed method.

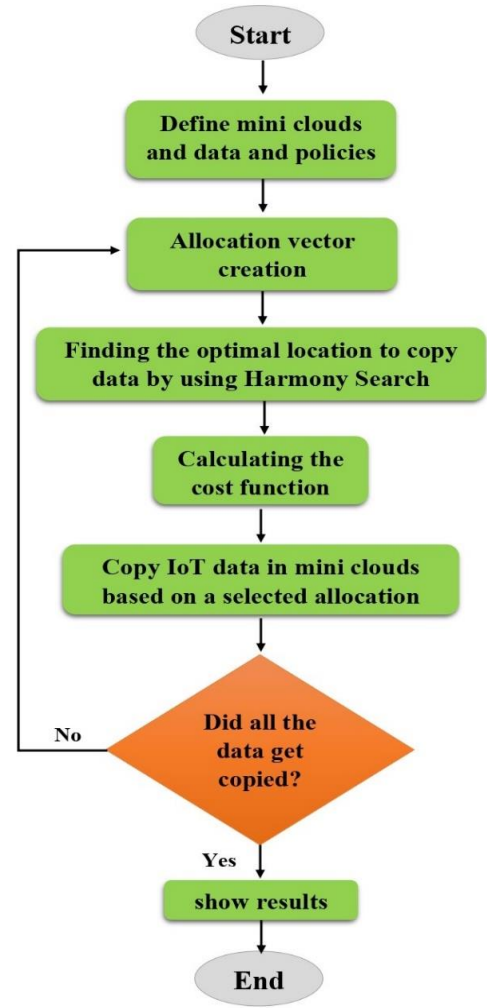

**Figure 2.** Flowchart for the proposed method.

### 4.1 HS-Based Allocation of Replicas

Figure 3 indicates another schematic view of the entire system, which consists of three sections: gateways, mini clouds,

and policies. A policy states which data should be replicated and for how many times, whereas a mini cloud includes four system specifications, *i.e.,* writing delay, reading delay, existing capacity, and total capacity, the values of which must be specified in every mini cloud. Moreover, a gateway has two sections, *i.e.,* gateway number and the data existing in the gateway. Finally, the existing data include two subsections, *i.e.* data number and data size.

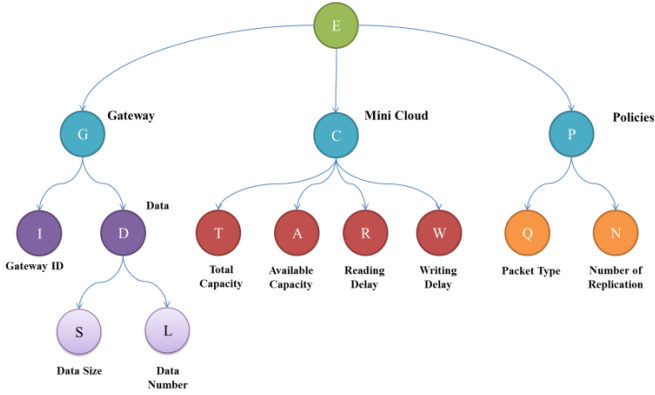

**Figure 3.** The total model of the proposed system.

The replica allocation process is performed for every datum pertaining to an object existing in the system gateway and therefore, a vector is used to allocate replicas. Every vector can be initialized from one to as many mini clouds as the number of replicas, elements, and the number of figures inside the elements. Note that the inside figures may not be duplicates. Therefore, the figure inside every element states the number of clouds into which data should be replicated. Consider Figure 4 for a more accurate description. In this case, Object 1 and Object 2 are assumed to be replicated two and three times, respectively. Hence, the array length will be different for each object, and the figures inside each array can be initialized with the number of mini clouds. To be accurate, three replicas of data are copied into Mini Clouds 2, 3, and 1 for Object 2.

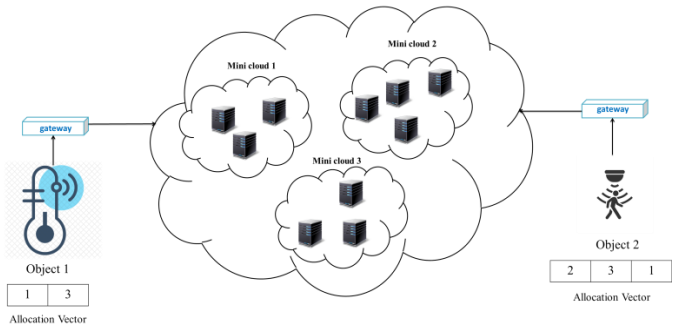

**Figure 4.** A case of allocation vector development.

The HS algorithm is responsible for finding an optimal vector (the best arrangement of figures) and allocating replicas accordingly to the mini clouds located in the cloud. The process of implementing the HS algorithm starts with defining every harmony as a sample solution/allocation vector in the problem. The main process is to select the best harmony or allocation vector. The HS algorithm is executed by generating initial random solutions (harmonies) as many as the sample population size. Each cell of harmony is called a musical instrument (*i.e.* the mini cloud number), and each value of an instrument is a tune.

The initial random harmonies are evaluated in terms of an aesthetic standard (cost function) to determine the cost of each harmony. After that, harmonies are sorted in an ascending order based on cost values. A few exercises are then performed, and a new harmony is generated in each exercise. New harmonies are generated by selecting two random harmonies through the roulette wheel algorithm. In the selection process, a harmony of lower-cost has a higher chance of selection. Two designated harmonies are then combined to generate a new harmony. Figure 5 indicates a case in which two harmonies are combined to generate a new harmony. In this case, a value is selected randomly from the two harmonies and placed into the corresponding cell in the new harmony.

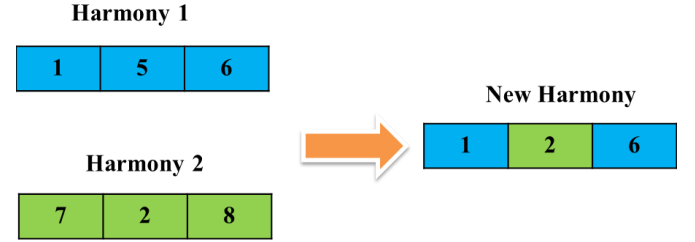

**Figure 5.** Combination of harmonies example.

In the next step, the cost function is calculated for the new harmony. After that, the cost of the new harmony is compared with that of the last harmony from the initial population. If the cost of the last harmony is higher, the new harmony replaces the current one. The population is then sorted into costs. This procedure continues until the maximum replication/exercise, at the end of which the most appropriate harmony with a lower cost is selected as the final solution or the final allocation vector. The IoT data are then copied into mini clouds based on the values existing in the designated vector. The cost function [30] (in which time is measured in seconds) is defined as (1) for every datum in the proposed approach [31]:

$$Min_{c=1,\ldots,n}\left\{T_{dj} + \left(\left(\frac{1}{B_{jc}} + R_j + W_c\right) * L_d\right) + Max_{c'\subset c}\left(T_{dc} + \left(\frac{1}{B'_{cc'}} + R_c + W_{c'}\right) * L_d\right)\right\} \quad (1)$$

In Equation 1, the first part pertains to data transfer from the gateway to the entire cloud: $T_{dj} + \left(\left(\frac{1}{B_{jc}} + R_j + W_c\right) * L_d\right)$ and the second part concerns data transfer between clouds: $Max_{c'\subset c}\left(T_{dc} + \left(\frac{1}{B'_{cc'}} + R_c + W_{c'}\right) * L_d\right)$

## 5. SIMULATION RESULTS

This section analyzes the results of implementing the proposed approach in MATLAB 2019. The experiment environment was considered a cloud computing system with a number of mini clouds and IoT devices. The proposed approach was compared with other algorithms (*e.g.* random search, genetic algorithm, and forest algorithm) in terms of different parameters. Moreover, four scenarios were defined to analyze the results accurately, for which simulations were performed on various scales. Table 2 and Table 3 show the parameters pertaining to the number of iterations and the parameters of reading/writing delay in gateways and mini clouds, respectively.

**Table 2.** Parameters of the number of iterations.

| Parameter | Value |
|---|---|
| **Timesteps** | 500 Second |
| **Minimum iterations** | 5 |
| **Maximum iterations** | 10 |
| **Minimum data size** | 20 bytes |
| **Maximum data size** | 100 bytes |

**Table 3.** Parameters of reading/writing delay for gateways and clouds in a millisecond.

| Parameter | Value |
|---|---|
| **Minimum writing delay** | 20 ms |
| **Maximum writing delay** | 70 ms |
| **Minimum reading delay** | 20 ms |
| **Maximum reading delay** | 70 ms |

Table 4 presents the values of different scenarios. The proposed algorithm was evaluated in four scenarios similar to those commonly used in the other papers of this field. In simulations, the system environment size, *i.e.,* the number of gateways and clouds existing in the system, is first determined. The system size was, therefore, considered on different scales in different scenarios.

**Table 4.** Parameters of different scenarios.

| Scenario | Number of Gateways | Number of Mini Clouds |
|---|---|---|
| **Scenario 1** | 22 | 8 |
| **Scenario 2** | 25 | 10 |
| **Scenario 3** | 32 | 15 |
| **Scenario 4** | 40 | 25 |

Figures 6-9 draw a comparison between the proposed approach and other methods in terms of the costs of different scenarios. The proposed approach achieves a lower cost than the other methods in all timesteps of different scenarios. Accordingly, the proposed method outperformed the other three methods (FOA: Forest Optimization Algorithm, GE: Genetic Algorithm, Random Algorithm) by requiring a lower cost of access to the replicated data.

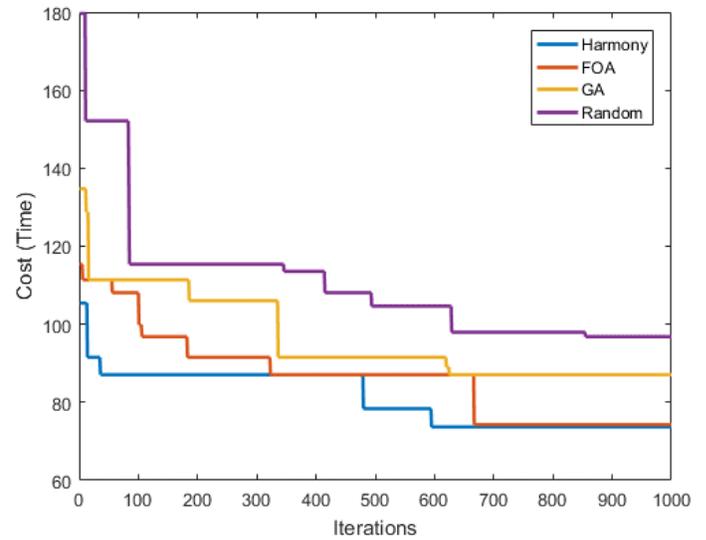

**Figure 6.** Comparing different scenarios in cost: Scenario 1.

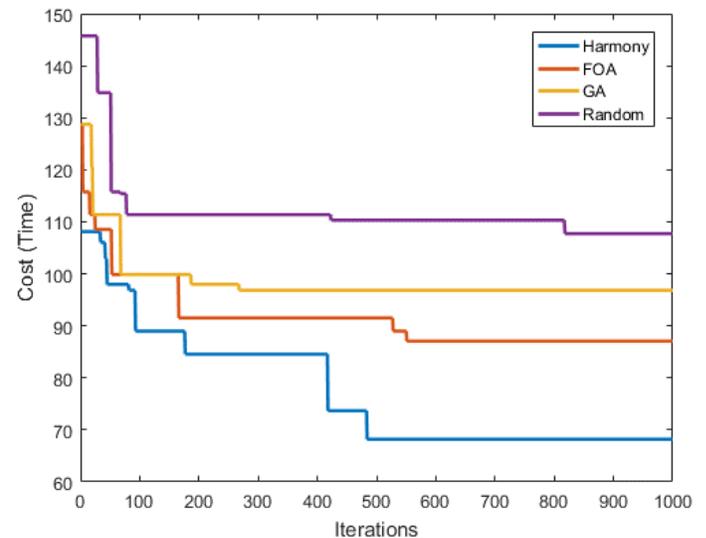

**Figure 7.** Comparing different scenarios in cost: Scenario 2.

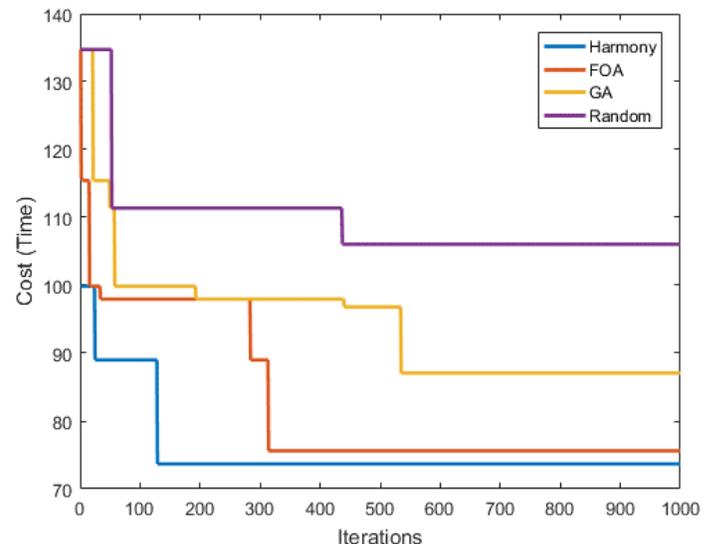

**Figure 8.** Comparing different scenarios in cost: Scenario 3.

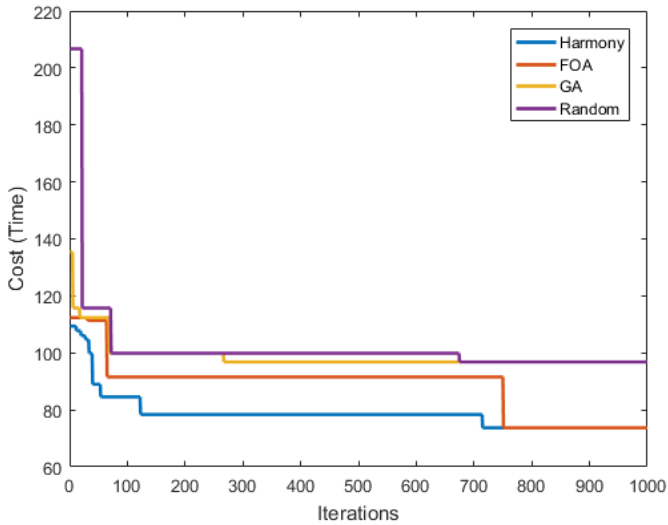

**Figure 9.** Comparing different scenarios in cost: Scenario 4.

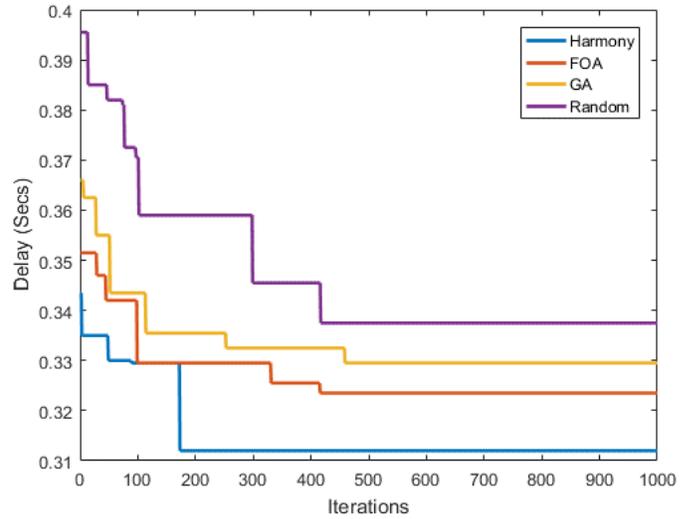

**Figure 11.** Comparing different scenarios in delay: Scenario 2.

Figures 10-13 draw a comparison between the proposed approach and other methods in terms of delay (access time) in different scenarios. According to the results, the proposed approach was characterized by shorter delays than the other methods in all scenarios. Selecting appropriate clouds and replicating data resulted in a more effective access and allowed achieving other goals such as reliability of replicas.

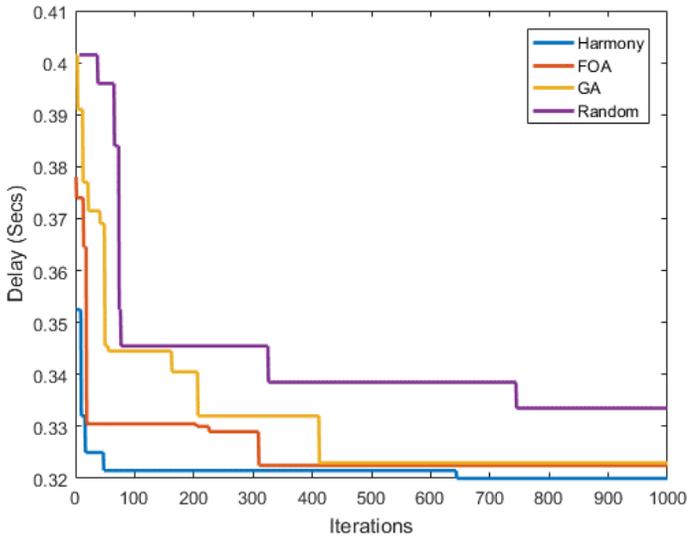

**Figure 10.** Comparing different scenarios in delay: Scenario 1.

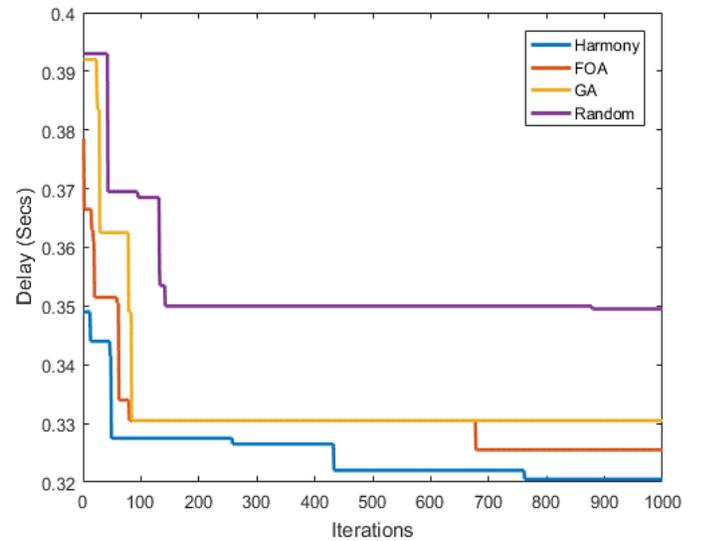

**Figure 12.** Comparing different scenarios in delay: Scenario 3.

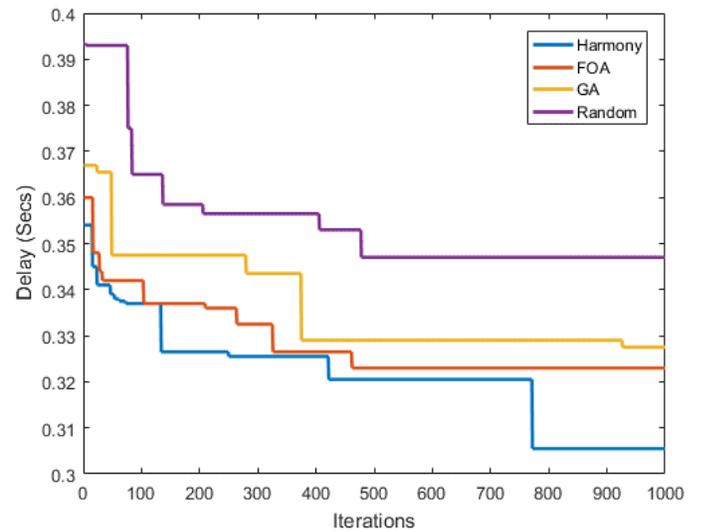

**Figure 13.** Comparing different scenarios in delay: Scenario 4.

Figures 14-17 make a comparison between the proposed approach and other methods in terms of energy in different scenarios. According to the results, the proposed approach consumed a lower amount of energy than the other methods in most of the timesteps and managed to allocate replicas in the cloud computing environment with a low rate of energy consumption.

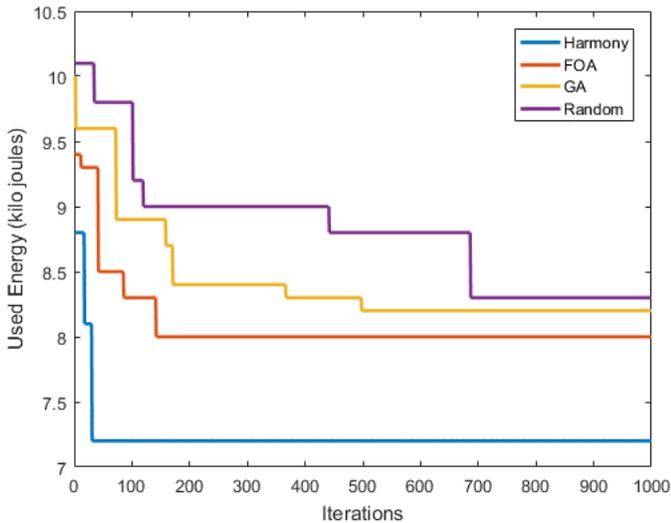

**Figure 14.** Comparing different scenarios in energy consumption: Scenario 1.

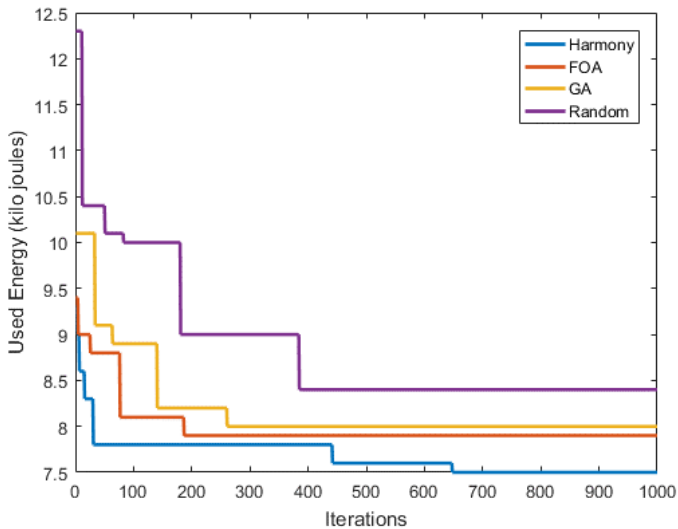

**Figure 15.** Comparing different scenarios in energy consumption: Scenario 2.

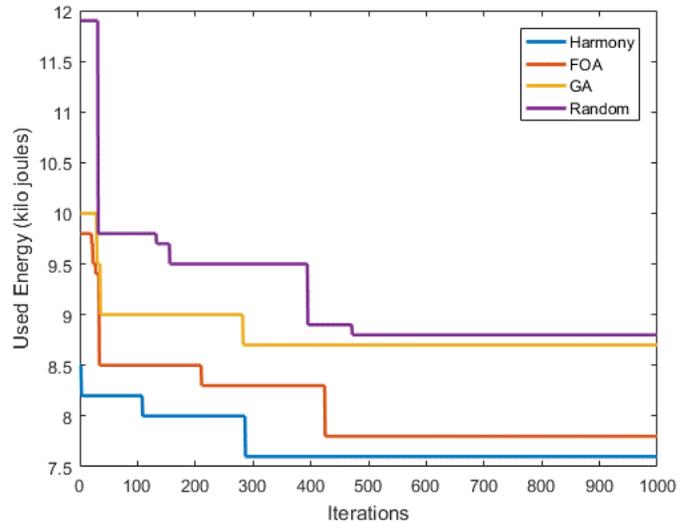

**Figure 16.** Comparing different scenarios in energy consumption: Scenario 3.

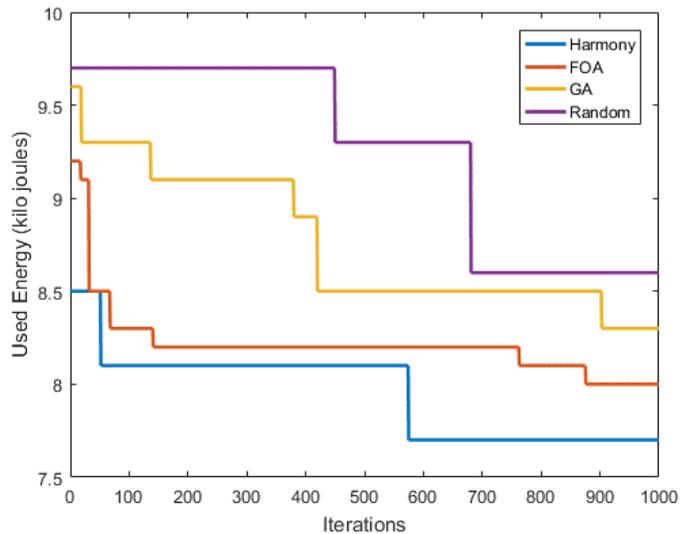

**Figure 17.** Comparing different scenarios in energy consumption: Scenario 4.

Figures 18-20 draw comparisons between the proposed approach and other methods (random search, genetic algorithm, and forest optimization) in different scenarios based on cost, delay, and energy consumption. The HS algorithm showed lower costs in the first and second scenarios than the other scenarios and the other algorithms. Regarding the delay, the third scenario showed a lower value than the other scenarios. However, the fourth scenario achieved a lower rate of energy consumption than the other scenarios. Therefore, analysis of different scenarios indicates that increasing the number of gateways and clouds resulted in better performance of the proposed approach than the other methods. Regarding data replication, the proposed approach guarantees the management of support replicas in terms of cost reduction, energy consumption, and increased access to data by generating various replicas.

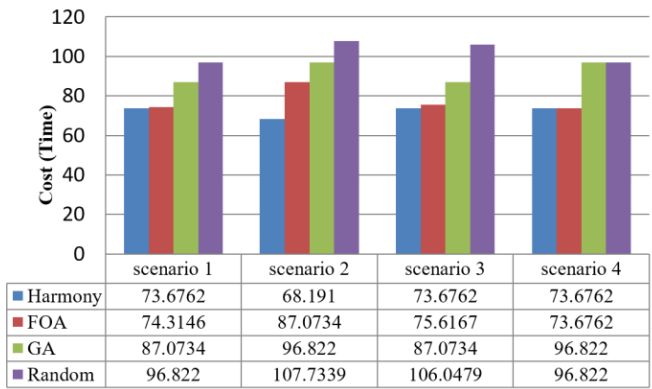

**Figure 18.** Comparing the proposed approach and other methods in cost function results.

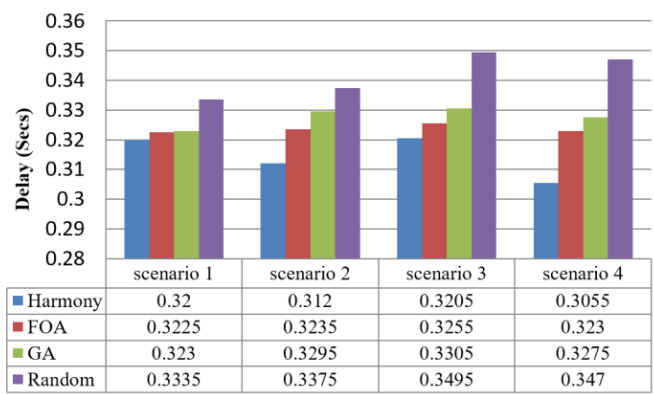

**Figure 19.** Comparing the proposed approach and other methods in delay results.

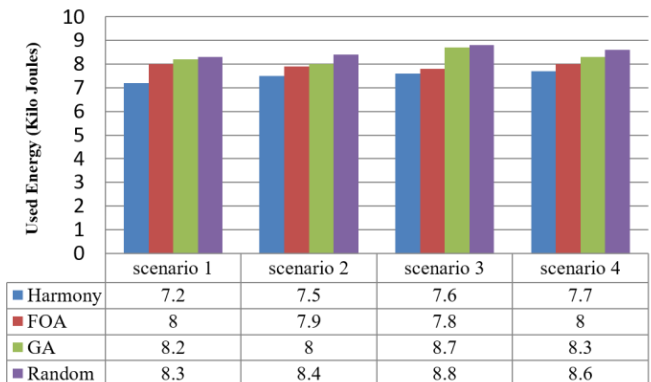

**Figure 20.** Comparing the proposed approach and other methods in energy consumption.

## 6. CONCLUSION

This paper proposes an approach based on the HS algorithm for allocation of replicas of IoT data in cloud computing environments. According to the simulation results of different scenarios, the proposed approach was able to allocate replicas faster and at a lower cost. In addition, the implementation results indicated that the proposed approach outperformed the other methods in terms of cost, delay, and energy consumption. Although many of the previous algorithms have mainly attempted to decrease cost and delay, the main goal may not always include optimization of the cost of replica allocation. It is necessary to consider other important parameters such as runtime, integrity of replicas, and capacity of storage resources. In addition, replicas can be allocated to the fog computing environment to decrease delay or congestion.

**Authors**

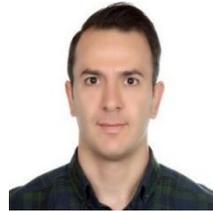

**Younes Jahandideh** received his Master's degree in Computer Engineering-Software in 2021 from Islamic Azad University of Ardabil Branch, Ardabil, Iran. His research interests are Cloud Computing, IoT, Wireless Networks, and Computer Communications (Networks).
Email: younes.jahandideh@yahoo.com

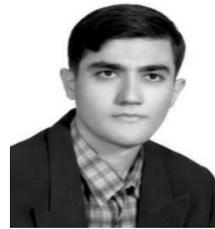

**Abbas Mirzaei** received the Master's degree in Telecommunication Engineering from Tabriz University, Tabriz, Iran, in 2007, and the Ph.D. degree in Computer and Communications from Malek Ashtar University of Technology. He currently works at the Faculty of Computer Engineering, Islamic Azad University, Ardabil Branch. His research interests include Computer Security and Reliability, Artificial Intelligence, Computer Communications (Networks), Cellular Network Optimization, Heterogeneous Networks, Data Mining, and Wireless Resource Allocation.

*Corresponding author. Email:
 abbas.mirzaei.1983s@gmail.com